\documentclass{PoS}
\usepackage{amsmath,amsfonts}
\usepackage[latin1]{inputenc}
\usepackage[T1]{fontenc}

\newcommand{\lb}{\langle \kern-.17em \langle} 
\newcommand{\rb}{\rangle \kern-.17em \rangle }

\newcommand{\beq}{\begin{eqnarray}}
\newcommand{\eeq}{\end{eqnarray}}

\newcommand\bs\boldsymbol

\title{The density of states approach to the sign problem}

\ShortTitle{Density of states}

     \author{\speaker{Biagio Lucini},$^a$ Olmo Francesconi,$^{b,c}$
     Markus Holzmann$^c$ and Antonio Rago$^d$\\
     \llap{$^a$} Mathematics Department, Computational Foundry,  College of
     Science, Swansea University, 
     Bay Campus, Fabian Way, Skewen SA1 8EN, UK\\
     \llap{$^b$}   Physics Department, College of Science, Swansea University,
     Singleton Campus, Swansea SA2 8PP, UK\\
     \llap{$^c$}   Univ. Grenoble Alpes, CNRS, LPMMC, 3800 Grenoble, France\\
     \llap{$^d$}
     Centre for Mathematical Sciences, University of Plymouth\\
     E-mail:  \email{b.lucini@swansea.ac.uk}, \email{o.francesconi.961603@swansea.ac.uk},
     \email{markus.holzmann@grenoble.cnrs.fr}, \email{antonio.rago@plymouth.ac.uk}
   }

   \abstract{
     Approaches to the sign problem based on the density of states
     have been recently revived by the introduction of the LLR
     algorithm, which allows us to compute the density of states
     itself with exponential error reduction. In this work, after a
     review of the generalities of the method, we show recent results
     for the Bose gas in four dimensions, focussing on the
     identification of possible systematic errors and on methods of
     controlling the bias they can introduce in the calculation.  
   }

\FullConference{XIII Quark Confinement and the Hadron Spectrum - Confinement2018\\
		31 July - 6 August 2018\\
		Maynooth University, Ireland}

\begin{document}

\section{Introduction}
Several relevant strongly coupled systems in Condensed Matter and Particle Physics are described by a complex action. Examples range from QCD at non-zero density to dense quantum matter and strongly correlated electron systems. In most of these cases, robust analytical approaches are not known and currently numerical methods provide the only {\em ab-initio} reliable tool of investigation.  

For the class of systems with a complex action, the partition function can be cast into the general form
\begin{equation}
\label{eq:zetacomplex}
	Z = \int [ D \phi ] e^{-\beta S_R [ \phi ]+ i \mu S_I[\phi]} \ , 
\end{equation}
where we have made explicit the decomposition of the action into its real part $S_R$ and imaginary part $S_I$, controlled respectively by the couplings $\beta$ and $\mu$. In the previous equation, $\phi$ represents the collection of quantum fields that describe the theory. 

When $\mu = 0$, Eq.~(\ref{eq:zetacomplex}) can be interpreted as a Boltzmann weight and standard Markov Chain Monte Carlo methods can be used in numerical studies of the corresponding system. Conversely, at $\mu \ne 0$, the path integral measure is complex and standard importance sampling methods are inadequate to generate an ensemble of representative configurations for the model. At the origin of this failure are the strong cancellations that arise between positive and negative contributions to the partition function, which leave us with a numerical result that is several orders of magnitude smaller than the positive and the negative parts of the integral. This cancellation is known in the literature as the {\em sign problem} (see~\cite{Gattringer:2016kco} for a recent review).

It is worth noting at this point that the sign problem may be related to our way of describing the system rather than to some of its intrinsic physical properties. Indeed, for some systems it is possible to rewrite the action using dual variables. In this dual formulation, the sign problem disappears and Monte Carlo methods are perfectly viable~\cite{Endres:2006xu,Gattringer:2012df}. Nevertheless, for several relevant systems (e.g. QCD at finite density), a dual formulation is not known. Hence, if we want to {\em solve the sign problem}, finding a new technique that is capable of handling the numerical cancellations in the direct formulation is paramount. While a single algorithm that enables us to successfully address all the systems with a sign problem can not possibly be provided, since this will amount to solve at least one non-polynomial complete problem in a polynomial time~\cite{Troyer:2004ge}, several recent attempts using various techniques (including Complex Langevin dynamics, dual formulation, analytic continuation, density of states and thimble methods, see~\cite{Gattringer:2016kco} a discussion) have shown a good degree of success in different models. 

Our contribution further develops the density of state calculation (originally proposed in~\cite{Gocksch:1988iz} and more recently discussed in~\cite{Anagnostopoulos:2001yb,Fodor:2007vv,Langfeld:2014nta,Gattringer:2015lra}) with the LLR algorithm~\cite{Langfeld:2012ah,Langfeld:2015fua}. Our proposal has two components: first, we determine numerically a positive-definite density of states to a very high precision, spanning around 20 orders of magnitude with approximately fixed relative error; then, we integrate analytically a smoothed interpolation of the latter. We shall use the interacting Bose gas in four dimensions as a case study to illustrate our approach. Numerical results for this model using the same algorithm have been presented in~\cite{Langfeld:2015qoa,Pellegrini:2015dkk}, where the main focus was mostly on the feasibility of performing the numerical integral. Other studies of density of states methods for complex action systems include~\cite{Gattringer:2015lra,Garron:2016noc,Giuliani:2016tlu,Garron:2017fta,Giuliani:2017fss}.

The rest of our work is structured as follows. In Sect.~\ref{sect:algorithm}, we review the generalities of the LLR algorithm for the determination of the density of states. Our numerical results are then presented in Sect.~\ref{sect:results}. Finally, we summarise our findings and discuss open directions in Sect.~\ref{sect:conclusions}. 
\section{The LLR algorithm}
\label{sect:algorithm}
Let us start for simplicity by considering an Euclidean Quantum Field Theory described by a real action $S$:
\begin{equation}
Z(\beta)=\int [ D \phi ] e^{-\beta S [ \phi ]} \ .
\end{equation}
The density of states, defined as
\begin{equation}
\rho(E)=\int [ D \phi ] \ \delta(S[\phi]-E) \ ,
\end{equation}
allows us to rewrite $Z$ as 
\begin{equation}
\nonumber
Z(\beta)=\int d E \ \rho(E) \ e^{-\beta E}= e^{- \beta F} \ ,
\end{equation} 
where the integral runs over all possible values $E$ of the action $S$ weighted by $\rho(E)$, which represents the density of numbers of configurations having $S = E$, and $F$ is the free energy of the system. In terms of $\rho(E)$ the expectation value of an observable $O(E)$ can be recast into the form
\begin{equation}
\nonumber
 \langle O \rangle=\frac{\int dE \ \rho(E) \ O(E) \ e^{-\beta E}}{\int
   dE \ \rho(E) \ e^{-\beta E}} \ .
\end{equation}
Hence, the numerical knowledge of $\rho(E)$ allows us to determine the expectation values of all observables that are function of $E$ and - in principle at least - to compute the free energy $F$, from which the thermodynamical or the relevant QFT properties of the system follow.  

The main issue affecting the numerical determination of the density of
states is the variation of the latter over several order of
magnitudes. The LLR algorithm, which has been inspired by the
successful Wang-Landau approach to systems with a discrete energy
spectrum~\cite{Wang:2000fzi}, allows us to obtain a
piecewise-continuous approximation of the logarithm of the density of
states that has a controlled and exponentially suppressed error. Both
these features are important for the correct reconstruction of the
density of states: the fact that the error is controlled means that
the method is a first-principle approach; having an exponentially
suppressed error, in turn, guarantees that the numerical effort does
not depend on the local value of the density of states, but only on
the degree of accuracy that one wants to reach. 

The LLR algorithm is implemented through the following steps~\cite{Langfeld:2012ah,Langfeld:2015fua}:  
\begin{enumerate}
\item divide the (continuum) energy interval in $N$ sub-intervals of
  amplitude $\delta_E$
\item for each interval, given its centre $E_n$, define 
\beq
\label{eq:logrho_local}
\log \overline{\rho}(E) = a_n \left(E - E_n - \delta_E/2\right) + c_n
\qquad \mathrm{for~} E_n - \delta_E/2 \le E \le E_n + \delta_E/2 
\eeq
\item obtain $a_n$ as the root of the stochastic equation
\begin{equation}
\langle \langle \Delta E \rangle \rangle_{a_n} = 0 \Rightarrow
\int_{E_n-\frac{\delta_E}{2}}^{E_n+\frac{\delta_E}{2}} \left(E - E_n -
  \delta_E/2 \right) \rho(E) e^{-a_n E} dE = 0
\end{equation}
using the Robbins-Monro iterative method
\beq
\lim_{m \to \infty} a^{(m)}_n  = a_n  \ , \qquad a^{(m+1)}_n=a^{(m)}_n
- \frac{\alpha}{m} \frac{\langle \langle \Delta E \rangle
  \rangle_{a^{(m)}_n}} {\langle \langle \Delta E ^2 \rangle \rangle_{a^{(m)}_n}} \ .
\eeq
At fixed $m$, one has Gaussian fluctuations of $a^{(m)}_n$ around $a_n$
\item Define
\beq
c_n =  \frac{\delta_E}{2} a_1 + \delta_E \sum_{k=2}^{n-1} a_k + \frac{\delta_E}{2} a_n \ ,
\eeq
which, together with the numerically determined $a_n$, specifies the local approximation~(\ref{eq:logrho_local}). 
\end{enumerate}
\begin{figure}
	\begin{center}
	\includegraphics[width=0.42\textwidth]{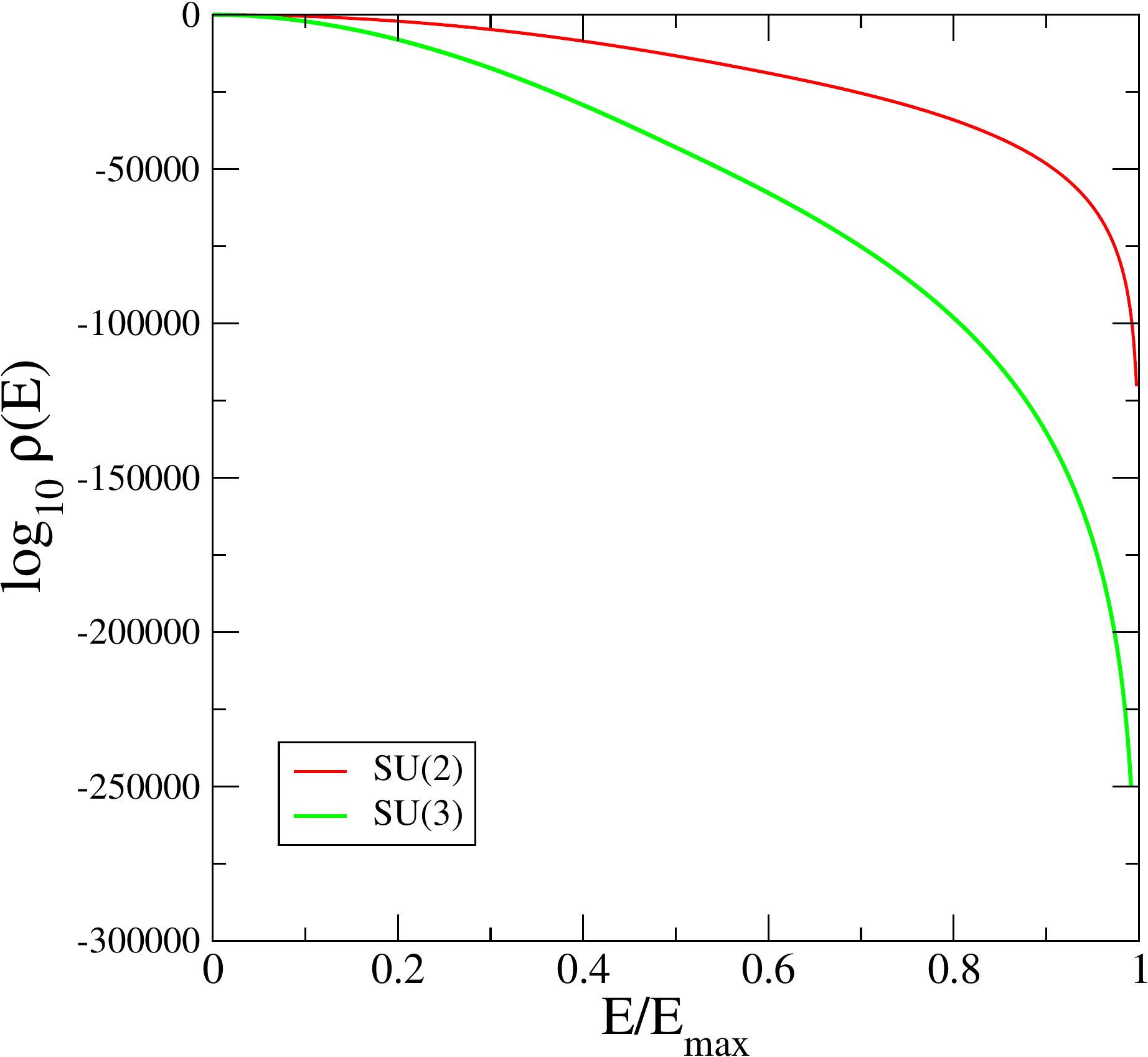} 
	~~
	\includegraphics[width=0.48\textwidth]{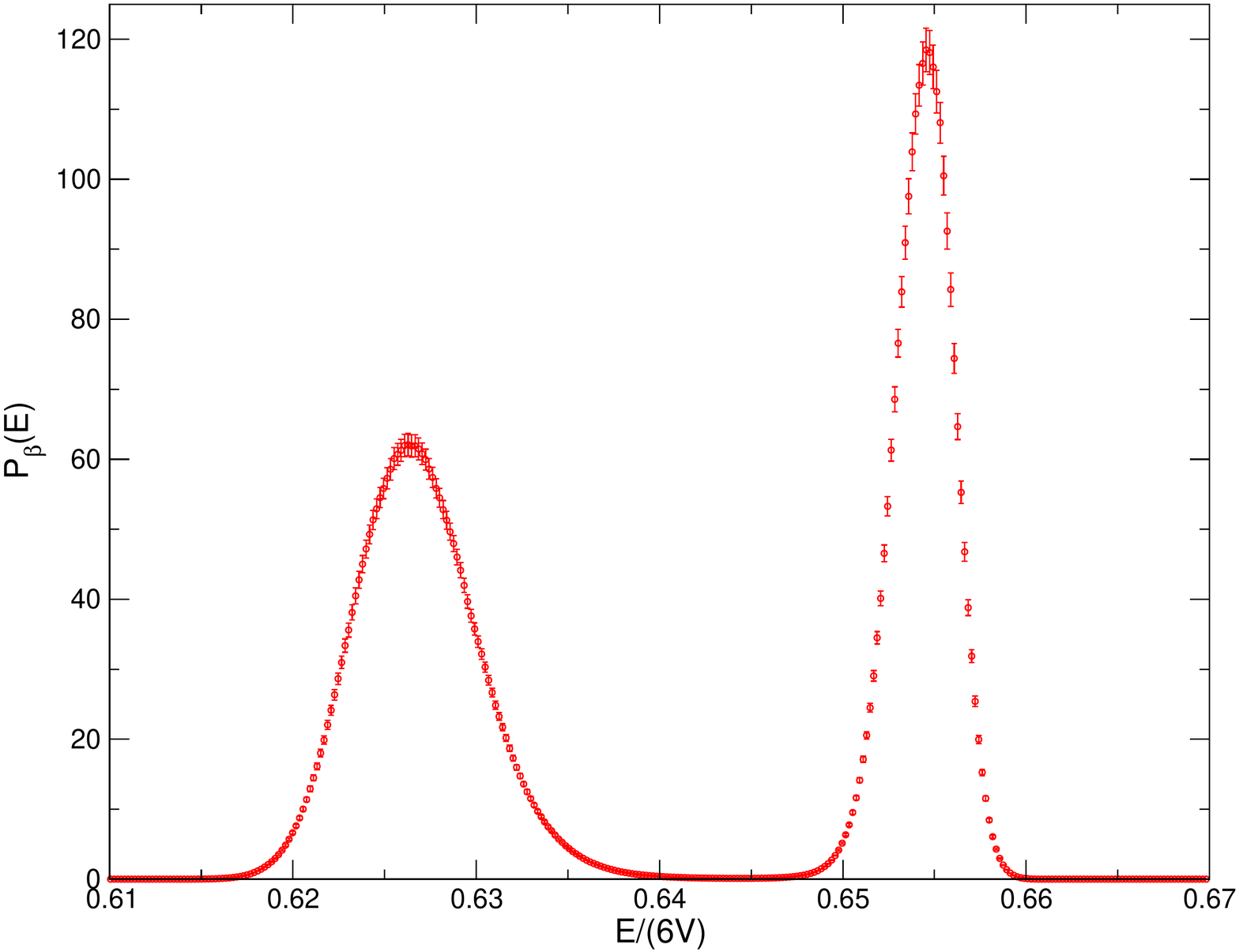} 	
	\end{center}
	\caption{Left: The density of states for the SU(2) and the SU(3) Lattice Gauge Theory models with the plaquette action. Right: Probability distribution at criticality for U(1) Lattice Gauge Theory on a $20^4$ lattice. \label{fig:density_su}}
\end{figure}
It is easy to see that, defining $\beta_{\mu}(E)$ the microcanonical 
temperature at fixed $E$, we have
\beq
\lim_{\delta_E \to 0} a_n = \left. \frac{\mathrm{d} \log \rho(E)}{\mathrm{d}
	E} \right|_{E = E_n} = \beta_{\mu}(E_n) \ .
\eeq
Under our assumption that $\rho$ is twice-differentiable, which holds
everywhere except for values of $E_n$ at which $\beta_{\mu}(E_n)$ corresponds
to a phase transition canonical $\beta$, away from the minimum of the
action $E_{min}$, $\overline{\rho}(E)$ converges quadratically to the density of
states $\rho(E)$ in the limit $\delta_E \to 0$. 

For ensemble averages of observables of the form $O(E)$, the
convergence to the expectation value computed with
$\overline{\rho}(E)$ to the canonical one is also quadratic in $\delta_E$,  
\beq
\langle \overline{O} \rangle_{\beta} = \frac{\int O(E) \ \overline{\rho} (E)
  \ e^{- \beta E} \ {\mathrm{d} E}}{\int \overline{\rho} (E) \ e^{- \beta E}
  {\mathrm{d} E}} = \langle O \rangle_{\beta} + {\cal O}\left(
  \delta_E^2 \right) \ .
\eeq
Moreover, we can prove that $\overline{\rho}(E)$ is measured with constant relative error (a feature that is known as {\em exponential error reduction}):
\beq
\frac{\Delta \overline{\rho}(E)}{\overline{\rho}(E)} \simeq \mathrm{constant} \ ,
\eeq
where $\Delta \overline{\rho}(E)$ denotes the statistical error on the
numerically reconstructed quantity $\overline{\rho}(E)$. 
To date, the LLR algorithm for real actions has been showcased in
several models. We show two applications in
Fig.~\ref{fig:density_su}, where in the left pane we report the
density for SU(2) and SU(3) and show that they can be accurately
determined respectively over 120000 and over 250000 orders of
magnitude~\cite{Langfeld:2012ah}. In Fig.~\ref{fig:density_su} (right), we show the
doubly peaked energy distribution at criticality in U(1) lattice gauge theory on a
$20^4$ lattice with periodic boundary conditions~\cite{Langfeld:2015fua}, which - owing to
severe metastabilities - is out of reach with traditional
importance sampling methods, including specialised ones. 

Controlled convergence to the exact value and exponential error
suppression are the two features of the algorithm that make it a
viable possibility for tackling the sign problem.
For the case $\mu \ne 0$ in Eq.~(\ref{eq:zetacomplex}), we define the generalised density of states as 
\begin{equation}
\rho(Q)=\int [ D \phi ]  \ e^{-\beta S_R [ \phi ]} \ \delta(S_I[\phi]-Q) \ , 
\end{equation}
in terms of which the partition function is obtained as
\begin{equation}
Z(\mu) = \int d Q \ \rho(Q) \ e^{i\mu  Q} \ .
\end{equation}
Due to the symmetry $\mu \to - \mu$, the partition function is real. However, the integrand is not positive definite. In fact, the integral proves to be strongly oscillating, with the oscillations giving rise to severe numerical cancellations. Therefore, in order to obtain a meaningful numerical result, $\rho(Q)$ needs to be known with an extraordinary precision. While the specific value of the latter depends on the problem at hand, at least a precision of order $10^{-20}$ on $\rho$ is in general necessary. The need to compute $\rho$ to such a high accuracy is the manifestation of the sign problem in the (generalised) density of states formulation. 
 
The severity of the sign problem is indicated by the vacuum
expectation value of the
phase factor in the phase quenched ensemble, which is defined by the action $S_R$, 
\begin{equation}
\langle e^{i\mu  Q} \rangle_{S_R} = \frac{Z(\mu)}{Z(0)} = e^{- V \Delta f } \ , 
\end{equation}
where $\Delta f$ is the specific free energy density difference between the original system and its phase quenched counterpart and $V$ is the total spacetime volume occupied by the system. In this language, the sign problem is an {\em overlap problem}. For future reference, we define the overlap free energy difference $\Delta F$ as
\begin{equation}
\Delta F = V \Delta f \ . 
\end{equation}
The motivation for using the LLR algorithm to compute $\rho(Q)$ mostly stems from the proven ability of this algorithm to solve overlap problems. 

However, one still needs to perform the integral with the required
accuracy, and for this the most direct approach (i.e. a numerical
Fourier transform of the piecewise approximation of the generalised
density of states) proves to be not accurate enough. The reason for
this failure is that one is bound to observe the singularities that
arise at points in which we connect the piecewise
approximations. These singularities have a frequency $1/\delta_Q$,
where $\delta_Q$ is the width of the interval for the restricted
sampling in Q. In addition, the data have a numerical error that
generates local fluctuations in $\log \rho$. Both these effects result
in a loss of precision that obfuscates the tiny signal modulated by
$\mu$. 

In order to bypass these difficulties, in~\cite{Langfeld:2014nta} it has been proposed to smooth the measured $a_k$ with a polynomial interpolation. More specifically, the smoothing consists in a compression of the generalised density of states using a global fit of the form
\beq
\log \rho(Q) = \sum_{i=0}^k \alpha_i Q^{2 i} \ .
\eeq
The effectiveness of the procedure has been demonstrated for the $\mathbb{Z}(3)$ spin model, which is the system which QCD reduces to 
at strong coupling, for large fermion mass and finite temperature. For finite $\mu$, the system is formulated as
\begin{eqnarray}
\nonumber
Z(\mu) = \sum _{\{\phi\}} \; \exp \Bigl\{ \tau \sum _{x,\nu } \left(
\phi_x \, \phi^\ast _{x+ \hat{\nu}} +  c.c. \right) + \sum_x \, \Bigl( \eta \phi_x + \bar{\eta } \phi^\ast _x
\Bigr) \Bigr\} = \sum _{\{\phi\}} \;\exp \Bigl\{ S_{\tau}[\phi] + S_{\eta}[\phi] \Bigr\} \ ,
\end{eqnarray}
where~$\phi_i \in \mathbb{Z}(3)$~is a spin variable defined on the sites $x$ of a three dimensional lattice of volume $V = L^3$, $\hat{\nu}$ is the unit vector in the direction $\nu$, $\tau$ is a coupling, $\eta = \kappa e^{\mu}$ and $\bar{\eta} = \kappa e^{- \mu}$, with $\kappa$ another coupling. The sum in the exponent is performed over all points $x$ and directions $\nu$, while the partition function is computed as a sum over all possible configurations $\{\phi\}$. We have explicitly separated the real part of the action (proportional to the coupling $\tau$) from the imaginary one (governed by $\eta$). We note that while the action is complex, the partition function is real.

\begin{figure}[htb]
	\begin{center}
		\includegraphics[width=.45\textwidth]{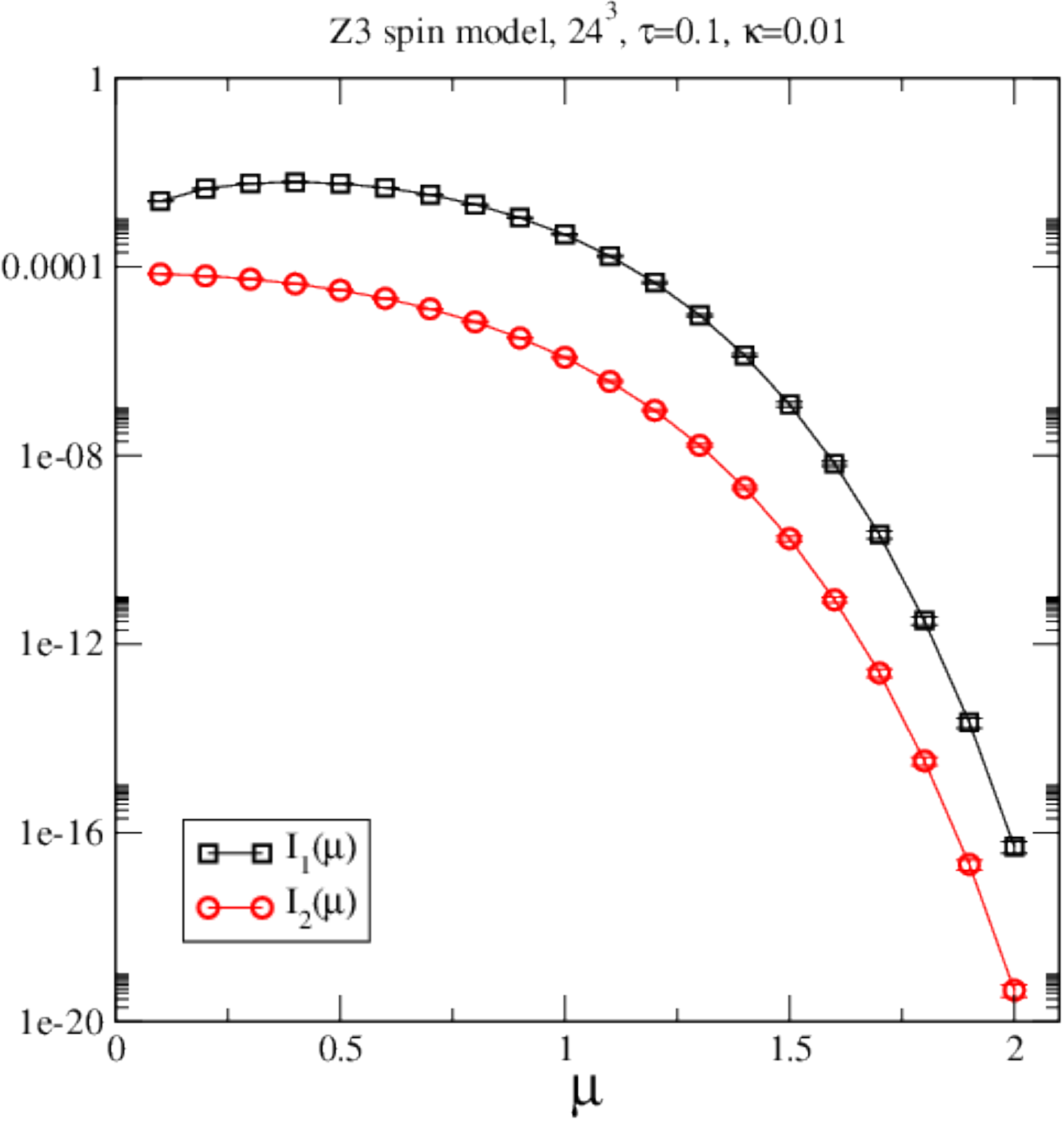}
		~~
		\includegraphics[width=.45\textwidth]{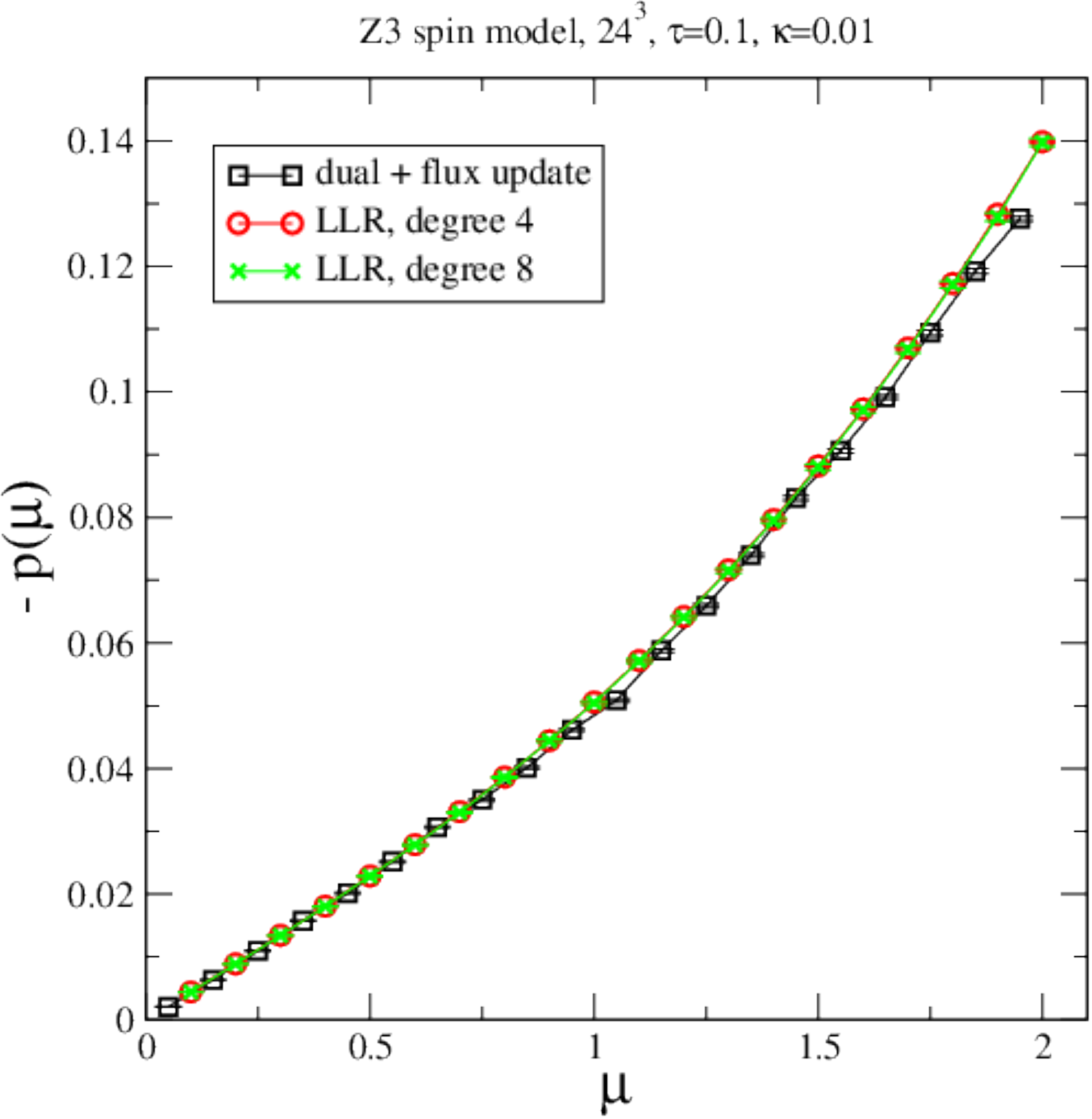}
	\end{center}
	\caption{Left: The numerator and the denominator defining the
          phase twist as a function of $\mu$. Right: the phase twist
          as a function of $\mu$. Both figures have been obtained on a
          $24^3$ lattice at $\tau = 0.1$ and $\kappa =
          0.01$. \label{fig:phasetwist}} 
\end{figure}
This model has been simulated using complex Langevin techniques and the
worm algorithm, the latter providing reference benchmarks for novel approaches. It has been shown~\cite{Mercado:2014dva} that an observable that is particularly sensitive to the sign problem is the phase twist $p(\mu)$, defined as
\begin{equation}
p(\mu ) \; = \; i \, \frac{ \sqrt{3} }{ V } \; \langle N_z -
N_{z^\ast} \rangle = \frac{1}{V} \frac{\sum_Q Q \  \rho(Q) \sin \left(
  k \sqrt{3} \sinh (\mu) Q \right)}{\sum_Q \rho(Q) \cos \left(
  k \sqrt{3} \sinh (\mu) Q \right)}  = \frac{1}{V} \frac{I_1(\mu)}{I_2(\mu)} \ ,  
\end{equation}
where $N_z$ and $N_z^{\star}$ are respectively the number of spins
pointing along $z$ and $z^{\ast}$, the two non-trivial elements of
$\mathbb{Z}(3)$. In Fig.~\ref{fig:phasetwist} (left) we show numerical
results for the numerator and the denominator determining the phase
twist. Those quantities vary over 15-16 orders of magnitude in the
simulated range of $\mu$. Their ratio, however, has much less
variation (Fig.~\ref{fig:phasetwist}, right). Hence, for an accurate
determination of the phase shift, a very precise measurements of $I_1$
and $I_2$ is required. Fig.~\ref{fig:phasetwist} reports the
determination of the phase shift with the LLR method using two
interpolations of the the generalised density of states, respectively
with a polynomium of order $4$ and a polynomium of order $8$, which
provide compatible results. In the same figure, we report also results
determined with a simulation of the dual model using a worm algorithm,
which are not affected by the sign problem. The agreement between this
latter set of data and the ones obtained with the LLR (shown
in~\cite{Langfeld:2014nta}, from which the figures have been borrowed)
is striking.  

We stress that in order to obtain those results a smoothing of the density of states has been crucial. We can interpret the polynomial interpolation as a Taylor expansion of $\log \rho$ that, for some reason that deserves to be further understood, has a convergence radius covering the whole range of interesting $\mu$. A number of open questions remain. Among them, if we assume that a polynomial interpolation of the data can be used over the range of $a_k$ that contribute to the integral, we would need to study the sensitiveness to the order of the polynomial, since {\em a priori} we do not have any guidance on the optimal order. More in detail, one would expect that a minimal order will be determined by the goodness of the fit, while a maximal order is imposed ultimately by the number of data points and before that by the maximum information they expose. The main objectives of this contribution is to devise a physically motivated method to put a meaningful upper bound on the maximal order of the fitting polynomium and to study the sensitiveness of the result with respect to the polynomial interpolation as a function of its order when the latter varies in the optimal range. This will allow us to get a handle on the convergence of the method.

\section{The Bose Gas}
\label{sect:results}
We pursue the programme illustrated in the previous section in the
self-interacting Bose Gas in four Euclidean dimensions and at finite
density. The model is described by the action 
	\begin{eqnarray}
	\nonumber
	S  &=& \sum_{i,a}\bigg[ \frac{1}{2}\left(
	2d+m^2\right) \phi_{a,i}^2
	+ \frac{\lambda}{4}\left(\phi_{a,i}^2\right)^2 
	- \sum_{\nu=1}^3 \phi_{a, i}\phi_{a, i+\hat \nu} 
	-\cosh(\mu)\,  \phi_{a, i}\phi_{a, i+\hat 4} \bigg]
	+  i \sinh(\mu)\, \sum_{i,a,b} \varepsilon_{ab}\phi_{a, i}\phi_{b, i+\hat 4}\\
	&=& S_R + i \sinh(\mu) S_I \ , 
	\end{eqnarray}
where $\lambda$ is the self-coupling, $\mu$ the chemical potential and
$m$ the mass of the bosons. The field has been decomposed into its
real part $\phi_1$ and its imaginary part $\phi_2$. 
The phase diagram (consisting in a low-density phase separated by a
phase transition from a high-density phase) has been mapped out
numerically through simulations of the sign-problem free dual theory
in~\cite{Gattringer:2012df}, which finds very good agreement with
mean-field theory calculations~\cite{Aarts:2009hn}.  

\begin{figure}[thb]
	\begin{center}
		\includegraphics[width=.45\textwidth]{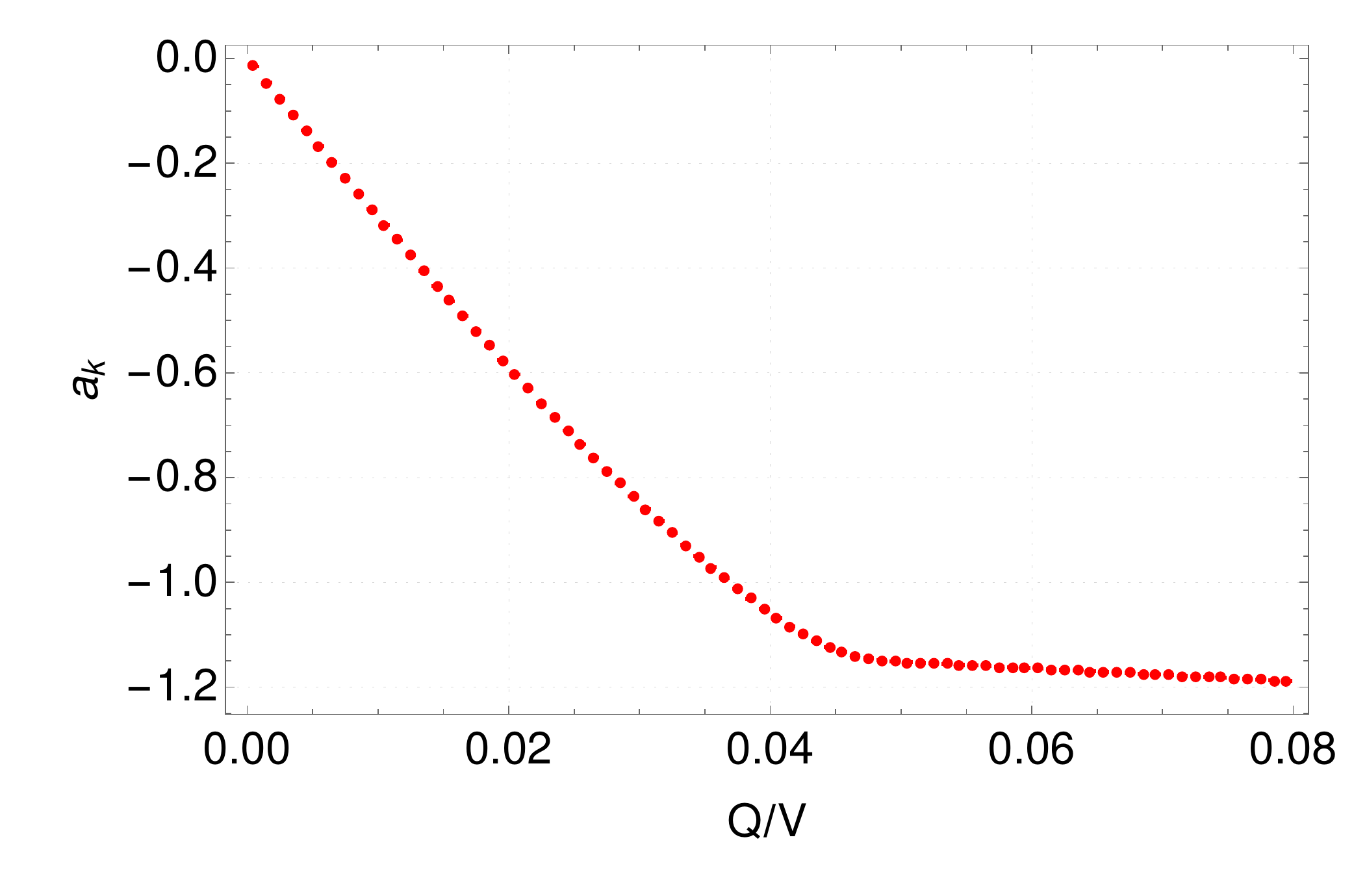}
		~~
		\includegraphics[width=.47\textwidth]{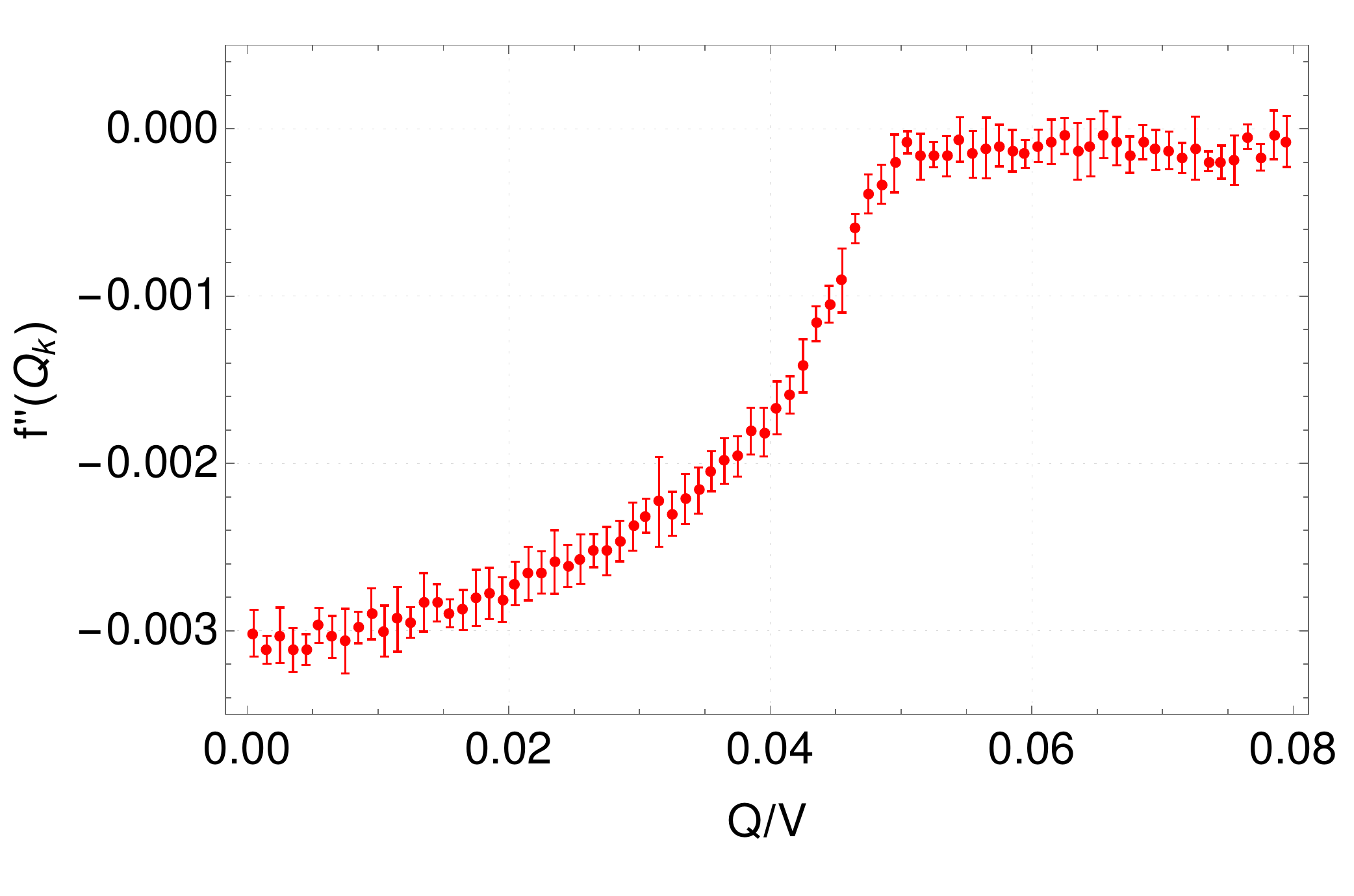}
		\caption{Coefficients $a_k$ (left) and their
                  derivative (right) for $\mu = 0.8$, $V = 10^4$.\label{fig:ak_d2}}
	\end{center}
\end{figure}

Throughout our calculation, we fix the self-interacting coupling $\lambda$ to the value $\lambda = 1.0$ and the particle mass $m$ to $m = 1.0$. We compute the density of states related to the imaginary action
\beq
S_I = \varepsilon_{ab}\phi_{a, i}\phi_{b, i+\hat 4} 
\eeq
performing a constrained Monte Carlo simulation for $S_R$. More in detail, we define 
\beq
\rho(Q) = \int \left[ {\cal D} \phi \right] \delta(S_I - Q) e^{- S_R} \ , 
\eeq
from which, using the LLR algorithm, we compute the quantities
\beq
a_k = \left. \frac{\mathrm{d} \log \rho}{\mathrm{d}Q} \right|_{Q_k} \ ,
\eeq
for chosen values of $Q_k$, which we take equally spaced. As an
example, we report in Fig.~\ref{fig:ak_d2} (left) the determination of
the $a_k$ for $\mu = 0.8$, $V = 10^4$.   
\begin{figure}[htb]
	\begin{center}
		\includegraphics[width=.45\textwidth]{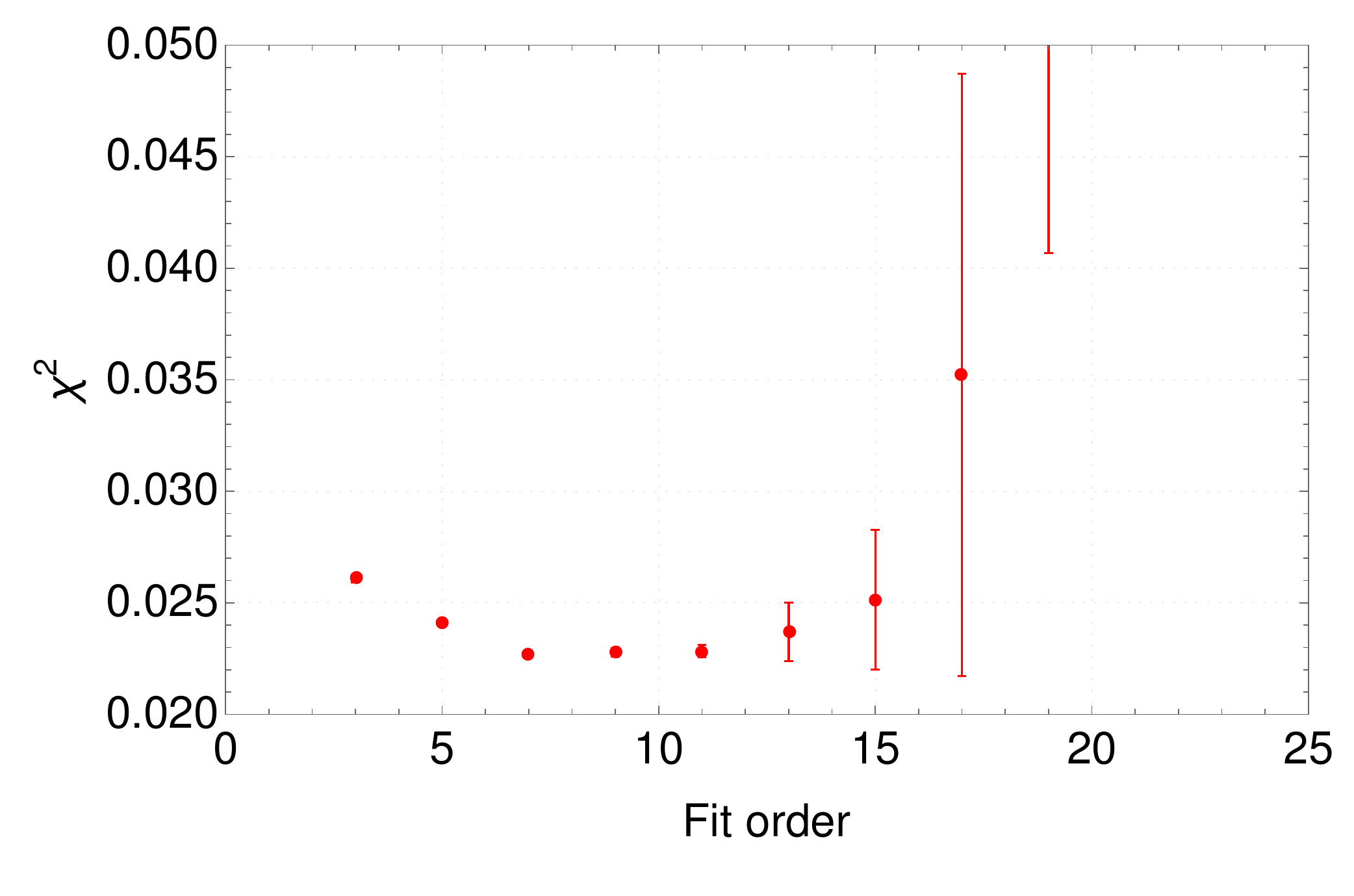}
		~~
		\includegraphics[width=.45\textwidth]{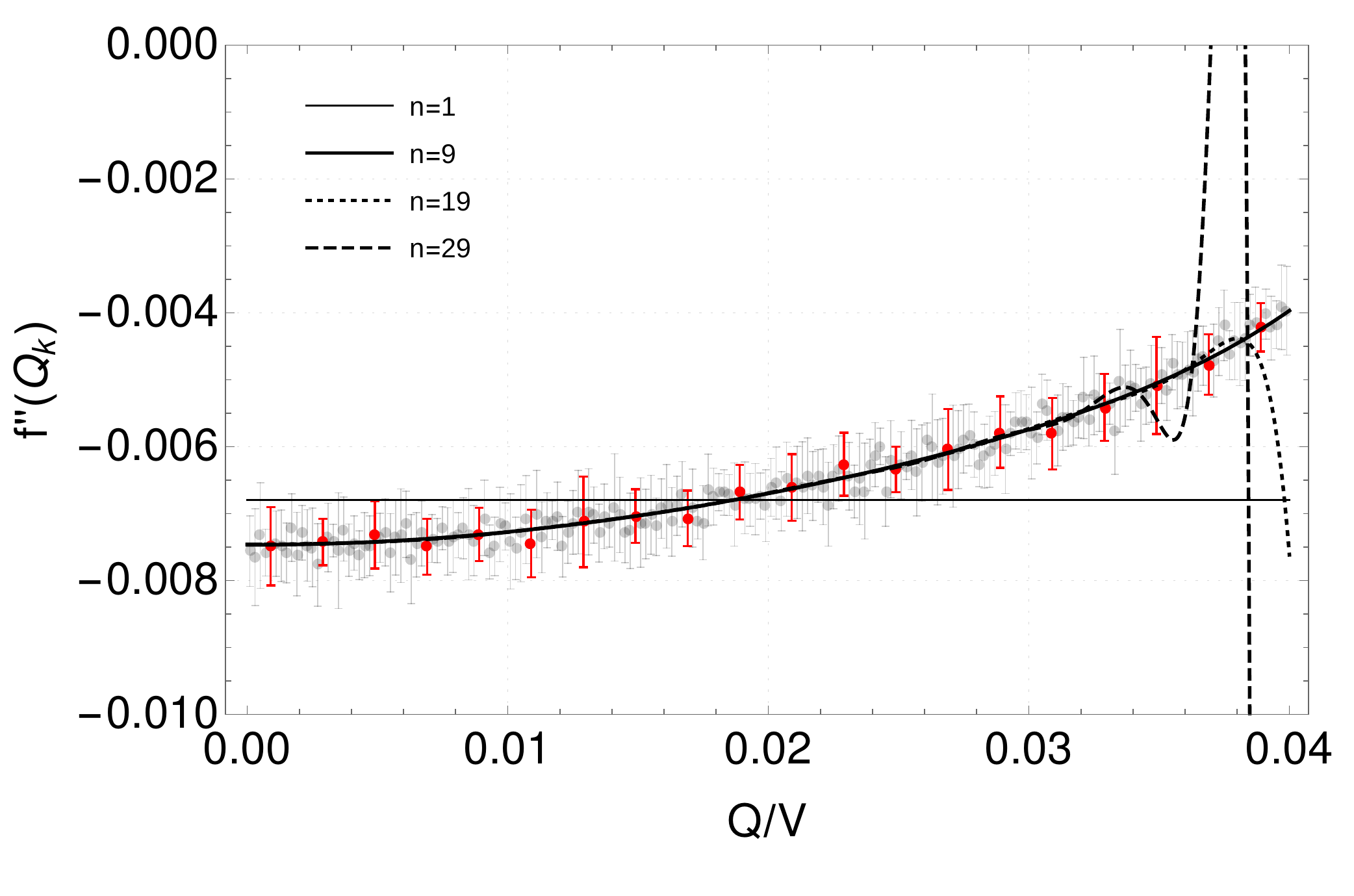}	
	\end{center}
	\caption{Left: Reduced $\chi^2$ for $f^{\prime
            \prime}$. Right: description of $f^{\prime \prime}$ using
          various order $n$ polynomial interpolations of the $a_k$ on a
          subset of the data (highlighted in red for the
          derivative). Both sets of results are obtained at $\mu =
          0.8$ and $V = 8^4$. \label{fig:testfit}}  
\end{figure}
As discussed previously, a piecewise approximation is not precise
enough to uncover the cancellations that typically take place in this
system. Hence, we resort to a polynomial interpolation over the whole
interval. This immediately opens the problem of the stability of the
polynomial fit with the order of the polynomium. If the functional
form we choose is not adequate to describe the data, the $\chi^2$ will
expose its failure. However, it is easy to see that one can improve
the goodness of the fit by increasing the order of the polynomium. The
latter process will result in a different type of failure, which is
now common to refer to as {\em overfitting}: if the number of
parameters is large enough, the fitted functional form will not
describe correctly the data despite the low $\chi^2$. This is
generally visible through unwanted oscillation of the
interpolation between consecutive data points. In our case, it becomes
paramount to detect even the slightest hint of overfitting, since any
oscillation, no matter how small, can affect the precision of the
cancellation. In order to better constrain the fit, we resort to the
second derivative of $\log \rho$, which is formally given by  
\begin{equation}
f^{\prime \prime} = \left. \frac{\mathrm{d}^2 \log \rho} {\mathrm{d} Q^2} \right| _{S_I, k} =
\frac{360}{\delta_Q^4}\left( s_2 - \frac{\delta_Q^2}{12} \right) +
\mathcal{O}(\delta_Q^2) \ , 
\end{equation}
with $s_2$ the order two cumulant evaluated with an average restricted
to the $k$-th interval and $\delta_Q$ the width of each interval. An
example determination of $f^{\prime \prime}$ is provided in
Fig.~\ref{fig:ak_d2} (right). 

Rather than using the second derivative of $\log \rho$ with respect to
$Q$ directly in the fitting procedure, we look at how well the
polynomial fit of the $a_k$ describes this quantity. This gives us
both a visual (through oscillations) and a quantitative indication of
whether the chosen functional form is overfitting the
data. Fig.~\ref{fig:testfit} shows an example of our
procedure. As the order of the polynomium describing the $a_k$
increases, one can see that oscillations in its derivative are more
evident, especially for larger values of $Q$. We tale this as an
indication of overfitting. Looking at the reduced $\chi^2$ obtained
with the description of the $f^{\prime \prime}$ data through the
derivative of the polynomial smoothing of the $a_k$, we find that a
range of optimal polynomial degrees for the simultaneous description
of the $a_k$ and the $f^{\prime \prime}$ can be identified. For
instance, in the case $\mu = 0.8$ and $V = 8$ the range for the degree
$n$ of the fitting polynomial is generally between $5 \le n \le 15$
(see Fig.~\ref{fig:testfit}, left). While this has been illustrated on
a specific example, the procedure gives similar results for other sizes
and different values of $\mu$.
\begin{figure}[htb]
	\begin{center}
		\includegraphics[width=0.46\textwidth]{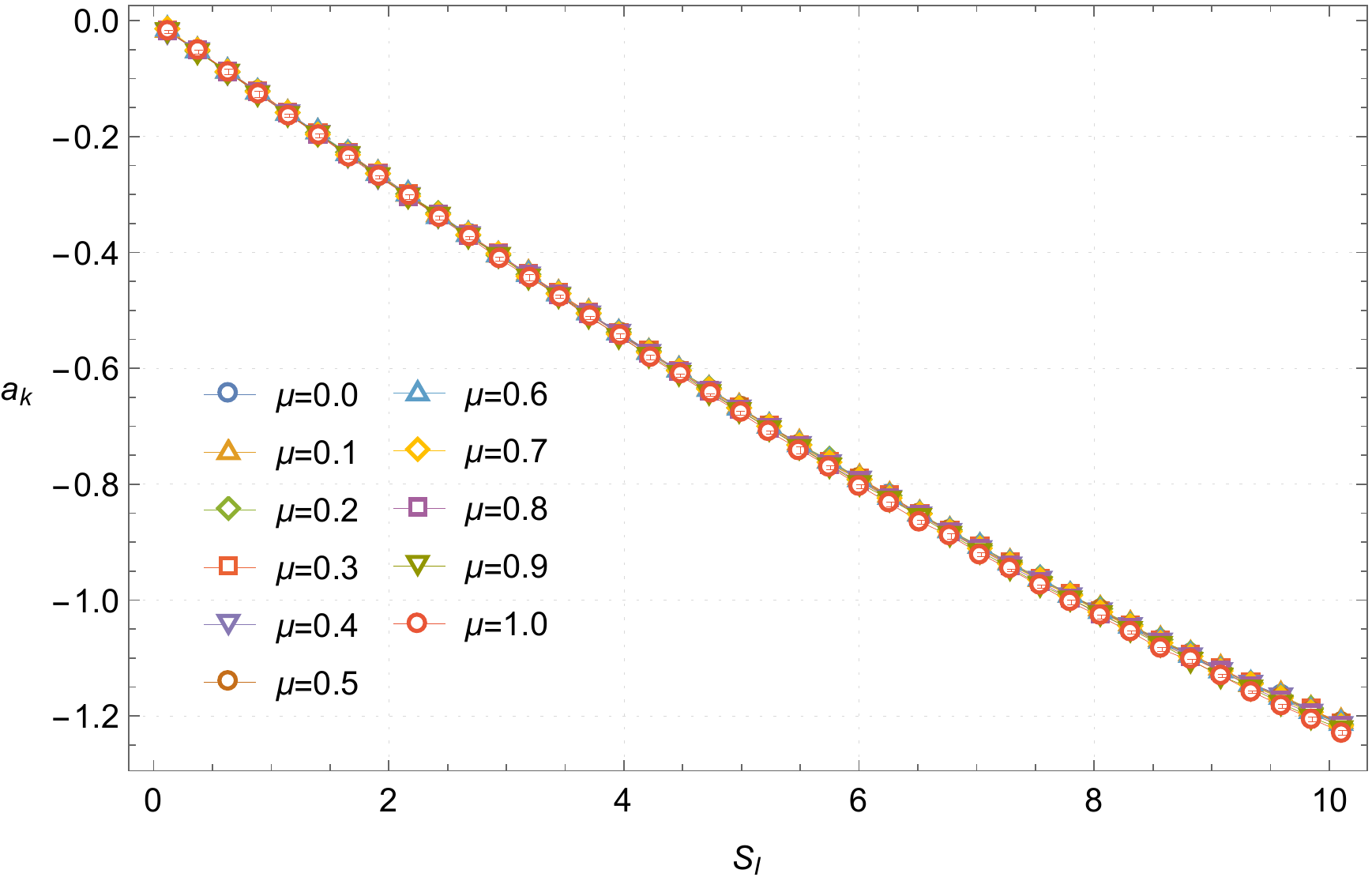}
		\includegraphics[width=0.48\textwidth]{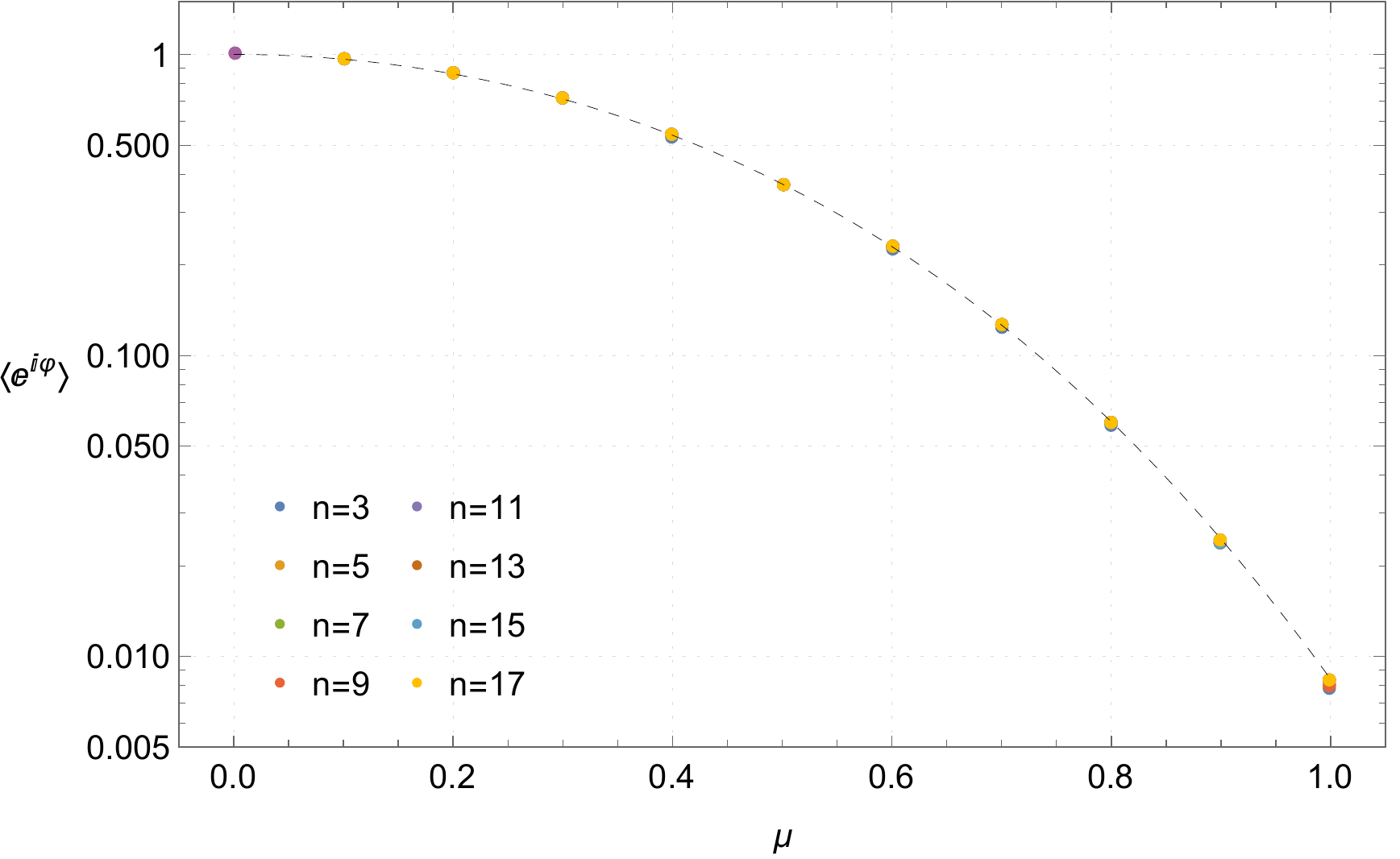}\\
		\includegraphics[width=0.46\textwidth]{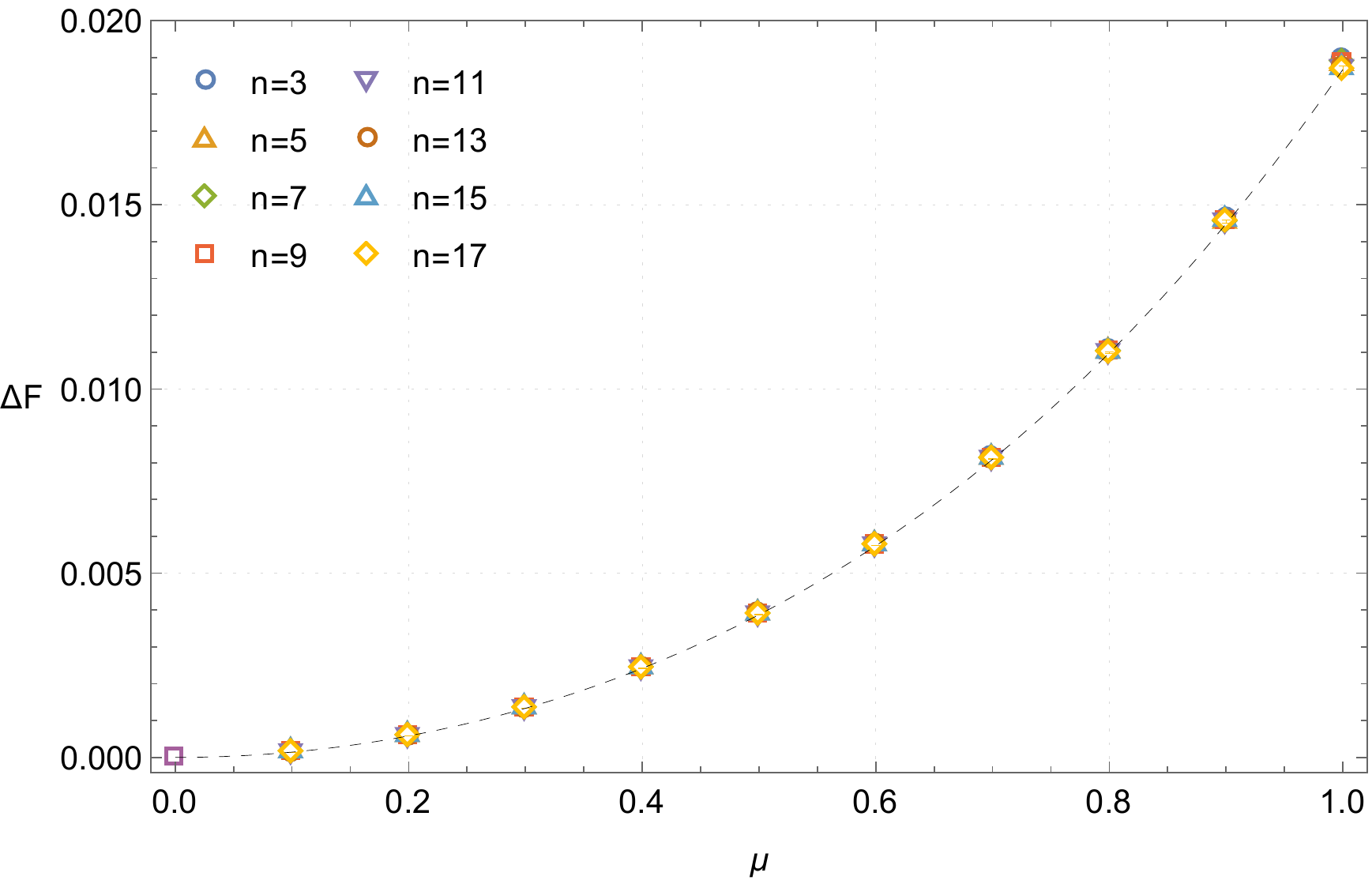}
		\includegraphics[width=0.49\textwidth]{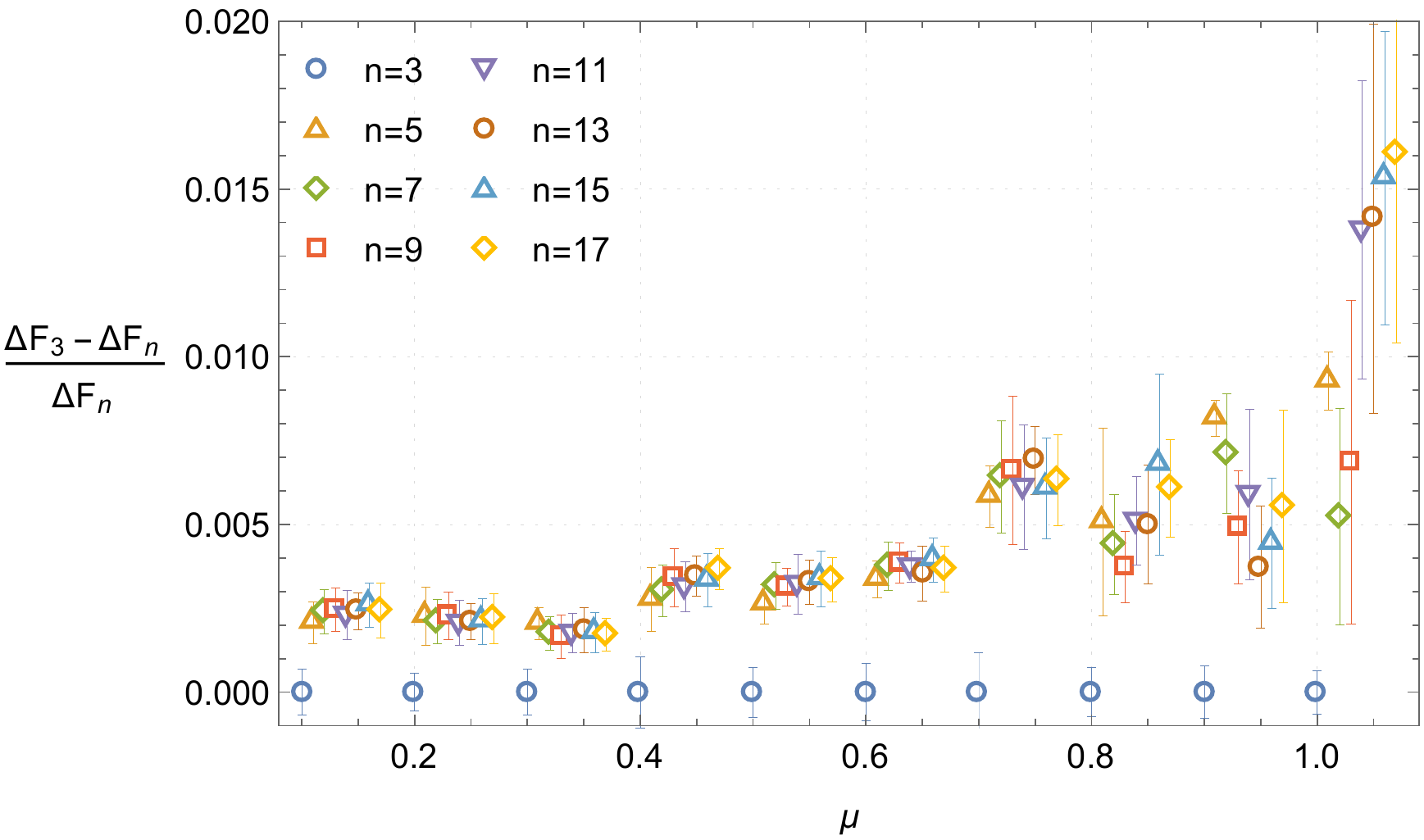}
	\end{center}
\caption{Top left: behaviour of the $a_k$ at the studied values of $\mu$ for an optimal choice of the fit order in each case. Top right: the phase average as a function of $\mu$ for polynomial fits of various orders in the optimal region. Bottom left: the extracted overlap free energy $\Delta F$ as a function of the fit order for optimal choices of the latter. Bottom right: convergence of the fit with the polynomial order, using as a reference the $n = 3$ result; note that for each $\mu$ the $n = 3$ result is positioned at the simulated value of $\mu$, while higher orders are displaced progressively on the right, for the sake of readability of the figure. All data shown are obtained on a $V = 4^4$ lattice. \label{fig:results_v4}}
\end{figure}

Having found a method for assessing the robustness of the fit, we now move to the determination of quantities of physical interest. First, we study the $a_k$ for a range of chemical potentials below the phase transition. The results for a $V = 4^4$ lattice are reported in Fig.~\ref{fig:results_v4}, top left panel. We see that for small values of $S_I = Q$ (with the displayed range being the one that contributes to the integral~(\ref{eq:zetacomplex})) the variation of these quantities is small as a function of $\mu$ on the scale of the figure up to the maximum studied value $\mu = 1.0$, which is relatively close to the critical value $\mu_c \simeq 1.15$~\cite{Gattringer:2012df}. Nevertheless, the reconstructed phase average (displayed in Fig.~\ref{fig:results_v4}, top right) varies over three orders of magnitude. From the phase average, we extract the overlap free energy $\Delta F$, which is shown in Fig.~\ref{fig:results_v4}, bottom left. The dependency of the latter on the fit order is plotted in Fig.~\ref{fig:results_v4}, bottom right. In this figure, we display the percentage variation of $\Delta F_n$, where the index $n$ refers to the order of the polynomial fit, with respect to the reference value $\Delta F_3$. For $\mu \le 0.8$ we see a clear plateau. At higher $\mu$, while the data are still compatible with a plateau, they display larger errors. At larger lattice sizes, the noise at those values of $\mu$ increases. However, we have found that the accuracy of the results can be still be reasonably controlled by a moderate increase of the accumulated statistics for the data. So far we have collected reliable results for $\mu \le 1.0$ and volumes up to $12^4$.  
\begin{figure}[thb]
	\begin{center}
		\includegraphics[width=0.55\textwidth]{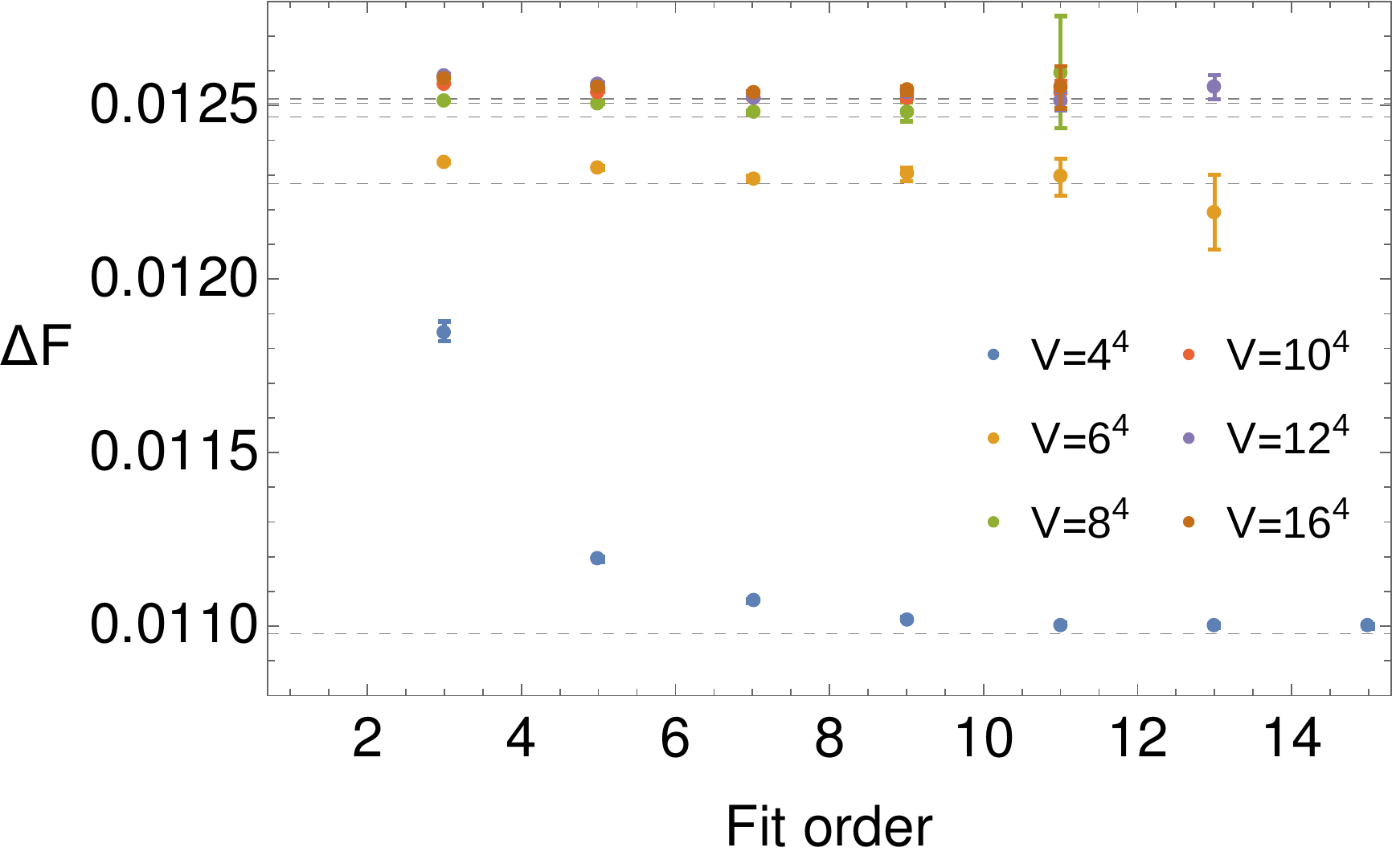}
		~~
		\includegraphics[width=0.39\textwidth]{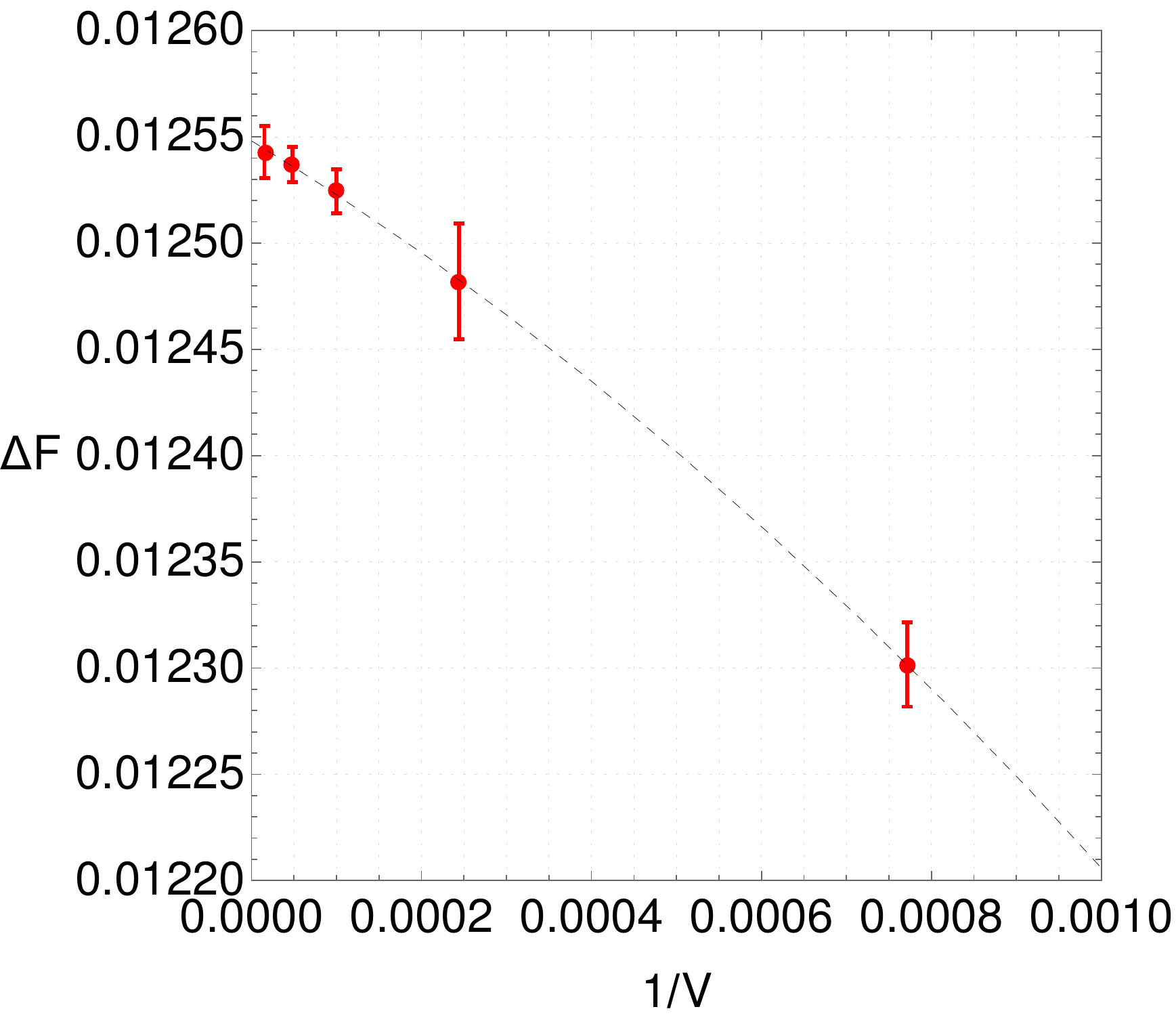}	
	\end{center}
\caption{Left: $\Delta F$ as a function of the polynomial fit order for the latter chosen in the optimal region at the shown values of the lattice volume. Right: extrapolation of the plateau value of $\Delta F$ to the infinite volume limit. \label{fig:ext_deltaf}}
\end{figure}

The quantity we are interested in here (which ultimately measures our
ability to extract numerical results with our approach) is the overlap
free energy in the thermodynamic limit. We show its determination as a
function of the fit order in Fig.~\ref{fig:ext_deltaf}, left; the data
show good convergence with the order of the polynomial used in the fit
at all simulated volumes.  We then take the plateau value and
extrapolate it to the infinite volume limit using a $1/V$ and a
$1/V^2$ correction, which appear to describe correctly our data (see
Fig.~\ref{fig:ext_deltaf}, right). This fitting ansatz provides us with the result 
	\begin{equation}
	\nonumber
	\Delta F = 0.012548(2)  - \frac{0.24(1) }{V} - \frac{98(17)}{V^2} \ ,
	\qquad \qquad 
	\end{equation}
whose relative difference from the mean field calculation $\Delta F_{MF}  \simeq 0.012522$~\cite{Aarts:2009hn} is of order $10^{-4}$. 
\section{Conclusions}
\label{sect:conclusions}
We have provided a numerical study of the self-interacting Bose gas using the density of states method. For the determination of the density of states, we have used the LLR algorithm, which has been proved to have significant advantages over traditional important sampling methods in cases in which one needs to measure exponentially suppressed signals and has been shown to be able to solve the sign problem for some toy models like the $\mathbb{Z}(3)$ spin model and heavy-dense QCD. With respect to applications involving a real action, in the complex action case an additional smoothing procedure of the density of states is needed. Here, we have provided a systematic study of this smoothing for the self-interacting Bose gas choosing as an interpolating function a polynomium of order $n$ and investigating the dependence of our results from $n$.  We have discussed criteria for assessing the robustness of the interpolation and shown that an optimal range of values of $n$ can be identified. Within this range, results appear to be independent of the chosen polynomial order. Using the developed methodology, for a particular choice of the chemical potential, we have provided an extrapolation to the infinite volume limit of the overlap free energy. The result we have obtained is compatible with mean-field, which has been shown to work well for this model. An extended calculations aimed to the determination of the infinite-volume $\Delta F$ is currently under way, and will be reported elsewhere. Our preliminary results indicate that with our technique we can determine the infinite volume value of $\Delta F$ up to $\mu = 1.0$, using finite volume results up to $V = 20^4$. We are currently investigating whether other improvements are needed for reaching higher $\mu$, closer to the critical value.

\acknowledgments
We thank L.~Bongiovanni, K.~Langfeld and R.~Pellegrini for
discussions. This work has been partially supported by the ANR project
ANR-15-IDEX-02. The work of BL is supported in part by the Royal
Society Wolfson Research Merit Award WM170010 and by the STFC Consolidated
Grant ST/P00055X/1. AR is supported by the STFC Consolidated Grant
ST/P000479/1. Numerical simulations have been performed on the Swansea
SUNBIRD system, provided by the Supercomputing Wales project, which is
part-funded by the European Regional Development Fund (ERDF) via Welsh
Government, and on the HPC facilities at the HPCC centre of the
University of Plymouth.  

\bibliographystyle{JHEP} 
\bibliography{conf18_lucini} 

\end{document}